\documentclass[
  aps,%
  12pt,%
  final,%
  notitlepage,%
  oneside,%
  twocolumn,%
  nobibnotes,%
  nofootinbib,%
  superscriptaddress,%
  noshowpacs,%
  centertags]%
          {revtex4}
\usepackage{amssymb}
\usepackage{amsmath}
\usepackage{graphicx}
\usepackage{psfrag}

\begin{document}

\title{ON THE STAR FORMATION RATE AND TURBULENT DISSIPATION IN THE GALACTIC
  MODELS}

\author{\firstname{E.~P.}~\surname{Kurbatov}}
\email{kurbatov@inasan.ru}
\affiliation{Institute of Astronomy, Moscow, Russia}

\begin{abstract}

We suggest a model for star formation function and a model for dissipation of
the turbulent energy of interstellar medium. Star formation function takes into
account the effect of turbulization of the ISM.  It is shown that application
of mentioned relations to the hierarchical scenario of formation of galaxies
allows to explain the observed delay of star formation in the Galaxy that
corresponds to the range of stellar ages from $8\--9$ to $10\--12$~Gyr.

\vspace{0.5cm}

\end{abstract}

\maketitle

\section{Introduction}

Observations of the galaxies reveal a great variety of physical processes
occurring in them.  However, it is possible to construct models of the galaxies
that contain  fairly small number of parameters.  For instance, in model~[1]
evolution of interstellar medium (ISM) is defined by the rate of supernovae
which depends on star formation history as well a by dissipation of turbulent
energy.

The models of galaxies with explicit modeling of ISM by gas-dynamical methods,
allow to resolve the regions with the scale not exceeding $10$~pc, while star
formation occurs over much smaller scales.  Complex structure of the ISM and a
wide range of temperature and density in the latter allow only phenomenological
approach to the description of star formation in the models that use explicit
modeling.  Over the time, several functions that determine star formation rate
(SFR) were suggested (see review~[2]). As an example we can list several
star-formation laws similar to the observations-based Schmidt law~[3] 
\begin{gather}
  \mathrm{SFR} \propto \rho^n  \;.
\end{gather}
The first example is the Kennicutt law~[4] which is often used in the numerical
models of galactic evolution~[5, 6]:
\begin{gather}
  \mathrm{SFR} \propto \frac{\rho}{\tau_{\mathrm{ff}}} \propto \rho^{3/2}  \;.
\end{gather}
The second example is dependence that is used in the models of evolutionary
synthesis~[2]:
\begin{gather}
  \mathrm{SFR} \propto {\mathrm{e}}^{-t/\tau_{\mathrm{sf}}}  \;.
\end{gather}
The third example is given by the model of star formation which is controlled
by ionization balance~[7]:
\begin{gather}
  \mathrm{SFR} \propto \rho^2  \;.
\end{gather}
As a shortcoming of Schmidt-type laws one may consider the fact that in these
models the rate of transformation of gas into stars depends on the local
density of the gas only and the effect of supernovae explosions  that enhance
the turbulent energy of ISM is lost; this results in decrease of the SFR.

Dissipation in the ISM is taken into account by consideration of shock
waves~[8], inclusion of artificial viscosity~[9, 10] and radiative
cooling~[11].  However, in the modeling of these effects the structure of the
ISM is not taken into account.

In Section~2 we analyze conditions for star formation in the galaxies, derive a
model for star formation and dissipation of the turbulent energy of the gas,
and describe a single-zone model for the evolution of galaxies and its
generalized version.   In Section~3 we present results of application of
suggested SFR-function to the single-zone model of the Galaxy.

\section{EVOLUTIONARY MODEL OF A GALAXY}

In this Section we construct a model for galactic evolution based on the
Tutukov-Firmani single-zone  model~[1, 12]. In the new model we use a more
accurate star formation function and apply a mechanism for dissipation of
turbulent energy that takes into account the structure of ISM; as well, we take
into account viscosity.

\subsection{Star formation rate}

Since one of the aims of the paper is derivation of the star formation model
for application in  the numerical models, it is reasonable to consider ISM over
the scale of the order of the minimum scale resolved in the numerical models
of the galaxies -- $10\--100$~pc, and over a time span that allows to consider
ISM as stationary.

Let consider SFR in the form
\begin{gather}
  \psi = c_* \frac{\rho}{\tau_{\mathrm{ff}}}  \;,
  \label{eq:schmidt_law}
\end{gather}
where $\rho$ is gas density, $\tau_{\mathrm{ff}}$ is the free-fall time; $c_*$
is so called dimensionless efficiency of star formation~[6].  The values of
this constant that are encountered in the galactic models vary from $0.1$~[13]
to $1$~[6].  Since it is deemed that the rate of star formation is proportional
to the amount of cold molecular gas~[14], it is reasonable to define the
dimensionless efficiency of star formation as a fraction of gas that is cold
and dense.

Let assume that the density perturbations in the gas are distributed in the
mass interval $(m_{\mathrm{min}}, m_{\mathrm{max}})$ and follow a power law,
i.e. $\mathbf{P}\{\mathrm{d}m\} \propto \mathrm{d}m^{1-\beta}$.  Then the
fraction of mass contained in all perturbations that have Jeans mass or higher
one, may be obtained by integration of mass spectrum over
$(m_{\mathrm{J}}, m_{\mathrm{max}})$ range:
\begin{gather}
  c_* = \frac{1 - (m_{\mathrm{J}}/m_{\mathrm{max}})^{\beta-1}}%
  {1 - (m_{\mathrm{min}}/m_{\mathrm{max}})^{\beta-1}}
  \left(\frac{m_{\mathrm{min}}}{m_{\mathrm{J}}}\right)^{\beta-1}  \;.
\end{gather}
If $m_{\mathrm{min}}$ and $m_{\mathrm{max}}$ are of the order of minimum
and maximum stellar mass respectively, i.e. $0.1\--100 M_{\odot}$, one may
assume $m_{\mathrm{min}}/m_{\mathrm{max}} \to 0$; then
\begin{gather}
  c_* \propto m_{\mathrm{J}}^{1-\beta}  \;.
\end{gather}
Inserting into Eq.~(5) expressions for Jeans mass
$m_{\mathrm{J}} \propto \rho^{-1/2} T^{3/2}$~[15] and free-fall time
$\tau_{\mathrm{ff}} \propto \rho^{-1/2}$~[6], one obtains: 
\begin{gather}
  \psi = g \rho^{\frac{1}{2}\beta + 1} T^{\frac{3}{2}(1 - \beta)}  \;,
  \label{eq:sfr_general}
\end{gather} 
where $g$ --- is a normalization constant (which will be defined below)
and $T$ is temperature.

It is necessary to clarify our notion of the temperature. Virial internal
energy of the interstellar gas per particle, averaged over all components of
the present-day ISM is approximately $4 \times 10^4 k_{\mathrm{B}}$ (where
$k_{\mathrm{B}}$ is Boltzmann constant).  This energy includes both thermal and
turbulent energy.  In the star formation region directly the temperature is of
the order of tens or several Kelvins~[16]. Since dependence~(10) is suggested
for inclusion in numerical  model, we have to proceed from the fact that at
hand there are only quantities averaged over the length scale of computational
cell of $10\--100$~pc, despite star formation occurs over much smaller length
scales.  However, assuming power-law for the mass spectrum of ISM components,
as above, as well as power-law dependence of dispersion of turbulent velocity
on the length scale~[17], one may claim that the average  the turbulent energy
of the ISM is proportional to the  turbulent energy and temperature over small
length scales, i.e., in the star formation regions.  Below, the temperature
$T$, will imply a quantity that is related to the maximum value of dispersion
of the turbulent velocity  $\sigma_0^2$ as
\begin{gather}
  \sigma_0^2 = \frac{k_{\mathrm{B}} T}{\mu}  \;,
  \label{eq:temperature}
\end{gather} 
where $\mu$ is the average molecular weight.

If the power of mass spectrum is Salpeter's one ($\beta = 2.35$),
$\psi = g \rho^{2.175} T^{-2.025}$.  Below we will assume for the ISM the
dependence
 \begin{gather}
  \psi = g \frac{\rho^2}{T^2}  \;.
  \label{eq:sfr}
\end{gather} 

We should note that quadratic dependence of SFR on density was assumed before
in the model of star formation governed by ionization~[7] and it recommended
itself well in the single-zone model of galactic evolution~[12]:
\begin{gather}
  \psi = f \rho^2  \;,\quad
  f = 2\times10^7 \text{~cm$^3$ g$^{-1}$ s$^{-1}$}  \;.
  \label{eq:sfr_tutukov}
\end{gather}
Proportionality constant $g$ may be found from  condition of the equality of
SFR in the models~(10) and (11) at virial temperature:
\begin{gather}
  g (4\times10^4)^{-2} = f  \;,
\end{gather}
which gives 
\begin{gather}
  g = 3.2\times10^{16} \text{~cm$^3$ K$^2$ g$^{-1}$ s$^{-1}$}  \;.
\end{gather}

Using Jeans criterion in the model for SFR, one may generalize the star
formation law to the case when account of the chemical composition of the gas
(its molecular weight), strength of magnetic field or rotation of a galaxy may
be important.

\subsection{Dissipation of turbulent energy}

Interstellar medium is a complex structure with energy, ionization, and
gas density varying over wide range.  One may note three approaches to the
description of the processes of formation and evolution of ISM:
``graviturbulent'', magnetohydrodynamical, and fractal.

The fractal approach is elaborated in~[18--20] and other studies (see
references  in~[21]).  This approach assumes that interstellar gas forms a
fractal structure.  Formation of molecular clouds in the framework of fractal
approach may be modeled as aggregation of mini-clumps of gas outflowing from
red giants~[21].

Within magnetohydrodynamical approach, the structure of ISM is governed by
magnetic field.  The key role in formation and evolution of giant molecular
clouds is accomplished by instabilities (thermal, Parker ones), while the
formation of the nuclei of molecular clouds is controlled by ambipolar
diffusion~[22--24].

In the graviturbulent model, ISM is a complex of random gas flows over the
length scale of the order of Galactic disk thickness.  The energy is
transferred from large scales to small ones; moreover, gas dynamics at large
length scales is defined by supersonic turbulence and over small scales it is
defined by gravitation~[24--25].

Magnetohydrodynamical and graviturbulent models are the most elaborated ones.
While all three models have their own advantages and shortcomings, we shall use
turbulent model as the base.

Our derivation of the rate of dissipation of turbulent energy in the galactic
gas will be based on the assumption that the turbulence exists within certain
range of length scales $(l_{\mathrm{min}}, l_{\mathrm{max}})$ and that
turbulent medium may be represented by an ensemble of the clouds with mass
distribution~[24]:
\begin{gather}
  \mathbf{P}\{\mathrm{d}M\} =
    \frac{\mathrm{d}M^{1 - \alpha}}{M_{\mathrm{max}}^{1 - \alpha}
      - M_{\mathrm{min}}^{1 - \alpha}}  \;,\quad
  \alpha \approx 1.5
  \label{eq:cloud_mass_spectrum}
\end{gather}  
and with power-law dependence of the average density of the cloud on its
size~[24]:
\begin{gather}
  \rho_l = \rho_0 \left( \frac{l}{l_{\mathrm{max}}} \right)^{-r}  \;,\quad
  r \approx 1.1  \;.
  \label{eq:cloud_density_spectrum}
\end{gather}
Let us also assume that the dependence of the dispersion of turbulent velocity
computed over a volume with length scale $l$ is a power-law function, as it is
confirmed by numerical models~[17, 26] and by observations~[16]:
\begin{gather}
  \sigma_l^2 = \sigma_0^2 \left( \frac{l}{l_{\mathrm{max}}} \right)^p  \;,\quad
  p \approx 1  \;.
  \label{eq:velocity_dispersion}
\end{gather}
In these formulas, the quantities $\rho_0$ and $\sigma_0$ may be considered as
averages over a volume with length scale $l_{\mathrm{max}}$.

Let assume that the dissipation of turbulent energy occurs via collisions of
clouds, their consequent compression by passing shock waves, and radiation of
thermal energy.  It is evident that the efficiency of radiation and, hence, the
efficiency of dissipation will be defined by the ratio of the duration of
collision and the time of radiation of thermal energy, i.e. cooling time~[27].
If it is assumed that the velocity of the size $l$ clouds is related to the
dispersion of velocities as $v_l^2 = \sigma_0^2 - \sigma_l^2$, the volume
density of turbulent energy associated with the clouds of the size in the range
$(l, l_\mathrm{d}l)$ is
\begin{gather}
  \rho_0 \frac{v_l^2 + \sigma_l^2}{2} \mathbf{P}\{\mathrm{d}l\} =
  \rho_0 \frac{\sigma_0^2}{2} \mathbf{P}\{\mathrm{d}l\}  \;,
\end{gather}
where $\mathbf{P}\{\mathrm{d}l\}$ is distribution of the clouds over the size.
The fraction of energy that is radiated during collision that lasts for
$\tau_{\mathrm{coll}, l}$, if the latter is small, is equal to the ratio of
collision duration and the cooling time:
\begin{gather}
  q_l = \tau_{\mathrm{coll}, l}
    \frac{\rho_{\mathrm{sh}, l}^2 \Lambda/\mu^2}{\rho_l \sigma_0^2/2}  \;,
\end{gather} 
where $\Lambda$ is cooling function per pair collision, $\mu$ is molecular
weight of the gas, $\rho_{\mathrm{sh}, l} = \xi_l \rho_l$ is the density of the
cloud after collision (after the passage of the shock wave), while coefficient
$\xi_l$ defines the density jump at shock wave.  Duration of collision
$\tau_{\mathrm{coll}, l} = l / D_l$ depends on the speed of the shock wave
$D_l$ which may be expressed via velocity of the cloud as $D_l = \eta_l v_l$.
In the general case, the rate of radiation of turbulent energy at all length
scales is equal to
\begin{gather}
  Q = \frac{\rho_0 \sigma_0^2}{2 \tau_{\mathrm{d}}}
    \int\limits_{l_{\mathrm{min}}}^{l_{\mathrm{max}}}
      \left(1 - \mathrm{e}^{-q_l}\right) \mathbf{P}\{\mathrm{d}l\}  \;,
  \label{eq:dissipation_rate_generic}
\end{gather}
where $\tau_{\mathrm{d}} = \sqrt{3/(2\pi G \rho_0)}$ is the timescale of
intervals between cloud collisions~[1, 28].

The spectrum of clouds dimensions $\mathbf{P}\{ {d}l\}$ may be found from the
following considerations.  Let represent the distribution of clouds over
mass~(14) as a combination of conditional probability
$\mathbf{P}\{\mathrm{d}M | l\}$ and distribution of the clouds over their size:
\begin{gather}
  \mathbf{P}\{\mathrm{d}M\} = \int_{l_{\mathrm{min}}}^{l_{\mathrm{max}}}
    \mathbf{P}\{\mathrm{d}M | l\} \mathbf{P}\{\mathrm{d}l\}  \;,
\end{gather}
and let write down the distribution of the clouds over mass as
\begin{gather}
  \mathbf{P}\{\mathrm{d}M | l\} = \delta(M - \rho_l l^3)\,\mathrm{d}M  \;.
  \label{eq:cloud_mass_with_size_distribution}
\end{gather}
Then, by virtue of~(14) and (15) we obtain
\begin{gather}
  \mathbf{P}\{\mathrm{d}l\} =
    \frac{\mathrm{d}l^{1 - \lambda}}%
      {l_{\mathrm{max}}^{1 - \lambda} - l_{\mathrm{min}}^{1 - \lambda}}  \;,
  \label{eq:cloud_size_spectrum}
\end{gather}
where $\lambda = (\alpha-1) (3-r) + 1 \approx 1.95$,
$M_{\mathrm{max}} = \rho_0 l_{\mathrm{max}}^3$, $M_{\mathrm{min}}
= M_{\mathrm{max}} (l_{\mathrm{min}}/l_{\mathrm{max}})^{3-r}$.

Thus, the expression for efficiency of dissipation $\epsilon_{\mathrm{d}}$ that
enters~(19)
as $Q = \epsilon_{\mathrm{d}} \cfrac{\rho_0 \sigma_0^2}{2 \tau_{\mathrm{d}}}$,
will be
\begin{multline}
  \label{eq:dissipation_efficiency_exact}
  \epsilon_{\mathrm{d}} =
    \frac{1-\lambda}{l_{\mathrm{max}}^{1-\lambda} - l_{\mathrm{min}}^{1-\lambda}}
    \int_{l_{\mathrm{min}}}^{l_{\mathrm{max}}} \mathrm{d}l\,l^{-\lambda}  \\
      \left\{ 1 - \exp\left[ -\frac{2 \Lambda l_{\mathrm{max}}}{\mu^2}
        \frac{\rho_0}{\sigma_0^3}
	\frac{\xi_l^2}{\eta_l}
	\frac{(l/l_{\mathrm{max}})^{1-r}}{\sqrt{1 - (l/l_{\mathrm{max}})^p}}
	  \right] \right\}  \;.
\end{multline}
For a mono-atomic gas the value of $\xi_l$ does not exceed~4, for diatomic one
it does not exceed~6~[29].  These limits are attained in strong shock waves
only.  The velocity of shock wave $D_l$ may exceed the velocity of colliding
clouds by one to one and a half orders of magnitude only, therefore we may
assume that approximately $2 \xi_l^2/\eta_l \approx 1$.  It appeared that the
spectrum of clouds sizes $\mathbf{P}\{\mathrm{d}l\}$ has a strong maximum at
$l = l_{\mathrm{min}}$; besides, $l_{\mathrm{min}}/l_{\mathrm{max}} \ll 1$ and
with sufficient accuracy we may use for the spectrum
$\mathbf{P}\{\mathrm{d}l\} = \delta(l - l_{\mathrm{min}})\,\mathrm{d}l$.
Finally we get
\begin{gather}
  \epsilon_{\mathrm{d}} =
    1 - \exp\left( -\frac{\Lambda l_{\mathrm{max}}}{\mu^2}
      \frac{\rho_0}{\sigma_0^3} \right)  \;.
  \label{eq:dissipation_efficiency}
\end{gather}
This function is obtained presuming that colliding clouds are of the same kind,
i.e., they have equal size.  If colliding clouds have different size, the
strong, exponential, dependence on arguments will relax.
 
Expression~(24) contains two important parameters: cooling function $\Lambda$
and maximum length scale of turbulence $l_{\mathrm{max}}$.  Cooling function
for the gas with the temperature $> 10^4$~K weakly depends on the latter and it
may be taken as $10^{-22}$ erg cm$^{-3}$ s$^{-1}$~[30].  The length scale of
turbulence was defined by trial and error method and it is equal to $5$~pc.
One may justify the order of magnitude of the latter scale by the fact that it
has to be limited by the size of stellar structures that form in the ISM, for
instance, open clusters which have typical length scale of several pc~[31].
Specific value of the maximum turbulence length scale was defined from the
condition of coincidence of the time span of observed and modeled interruption
of star formation process (see Sect.~3).

\subsection{Single-zone model}

In the Tutukov-Firmani single-zone model of galactic evolution~[1, 12] the mass
of the gaseous disk of the galaxy is defined by star formation process (which
has rate $\psi$), return of gas to the ISM by evolved stars, accretion of the
intergalactic gas ($\mathrm{d}M^{\mathrm{in}}/\mathrm{d}t$), and by the mass
loss via stellar wind, as well as by sweeping of the dust by stellar radiative
pressure and galactic wind ($\mathrm{d}M^{\mathrm{out}}/\mathrm{d}t$):
\begin{multline}
  \frac{\mathrm{d}M}{\mathrm{d}t} = -\psi +
    \int_{m_{\mathrm{min}}}^{m_{\mathrm{max}}}
      \mathrm{d}m\,\phi(m) [m - m_{\mathrm{r}}(m)]  \\
        \times \psi[t-\tau(m)]
      + \frac{\mathrm{d}M^{\mathrm{in}}}{\mathrm{d}t}
      - \frac{\mathrm{d}M^{\mathrm{out}}}{dt}  \;.
  \label{eq:mass_conservation}
\end{multline}
Equation for the energy balance takes into account energy input from supernovae
and dissipation of energy in clouds collisions:
\begin{gather}
  \frac{\mathrm{d}K}{\mathrm{d}t} + \frac{\mathrm{d}W}{\mathrm{d}t} =
    \epsilon_{\mathrm{SN}} E_{\mathrm{SN}} R_{\mathrm{SN}} -
    \frac{K}{\tau_{\mathrm{d}}}  \;,
  \label{eq:energy_conservation}
\end{gather}
where $K$ and $W$ are total kinetic energy and gravitational energy of the gas,
respectively; $\epsilon_{\mathrm{SN}}$ is the fraction of supernova energy
($E_{\mathrm{SN}}$) transferred to the gas per supernova (it is assumed in the
model that $\epsilon_{\mathrm{SN}} E_{\mathrm{SN}} = 0.05\times10^{51}$~erg),
\begin{multline}
  R_{\mathrm{SN}} = \frac{10^{-3}}{M_{\odot}}\,\psi(t-\tau_{\mathrm{SNI}})  \\
    + \int_{m_{\mathrm{low}}}^{m_{\mathrm{max}}}
    \mathrm{d}m\,\phi(m) \psi[t-\tau(m)]
  \label{eq:sn_rate}
\end{multline}
is the rate of supernovae, $\tau_{\mathrm{SNI}} = 10^9$~yr is the delay of type
Ia supernovae~[32],
\begin{gather}
  \tau_{\mathrm{d}} = \sqrt{\frac{2}{3\pi G \rho}}
  \label{eq:dissipation_time}
\end{gather}
is the timescale of dissipation of the turbulent energy in clouds
collisions~[1, 28].

As expression for gravitational energy, Tutukov and Firmani~[1] suggested to
use
\begin{gather} 
  W = \frac{G M_{\mathrm{G}} M H^2}{R^2 H_{\mathrm{s}}}  \;.
  \label{eq:grav_energy}
\end{gather}
Here $M_{\mathrm{G}}$ is the total mass of the galaxy, $H_{\mathrm{s}}$ is the
thickness of the galactic disk.  In this model it is assumed that the galaxy is
permanently in the state of virial equilibrium:
\begin{gather}
  K = \frac{3}{2} |W|  .
  \label{eq:virial_relation}
\end{gather}
The virial relation allows to obtain an equation for the thickness of the disk:
\begin{gather}
  \frac{\mathrm{d}H}{\mathrm{d}t} =
    \frac{\epsilon_{\mathrm{SN}} E_{\mathrm{SN}} R_{\mathrm{SN}} R^2
    H_{\mathrm{s}}}{5 G M_{\mathrm{G}} M H}
    - \frac{3 H}{10 \tau_{\mathrm{d}}}  \;.
  \label{eq:half_thickness_old}
\end{gather}
A consequence of the condition of virial equilibrium is that the state of the
system is defined by two quantities only -- by the mass and thickness of the
gaseous disk.

In the model, abundances of different chemical species are computed too.  For a
detailed description of the model see~[12].  The model has a minimum set of
free parameters and allows to follow all integral parameters of the galaxies,
such as SFR, luminosity, metallicity.  According to the computations carried
out in ~[12, 33--35], Tutukov-Firmani model reproduces correctly the star
formation history, the history of chemical enrichment of the galaxies and of
the intergalactic medium.

A peculiarity of Tutukov-Firmani model is that the density of gas and its
kinetic energy are not independent parameters, by virtue of virial
relation~(30).  It is easy to modify the model in order to remove this
peculiarity.  With this aim, one may introduce an additional variable --
turbulent energy of the gas $U$ and to use Newton [36] equations to describe
the motion of the disk boundary. In the Newton equation, in addition to
gravitation and pressure caused by inner gas energy, it is necessary to take
into account gas viscosity:
\begin{gather}
  \frac{\mathrm{d}H}{\mathrm{d}t} = V  \;,
  \label{eq:half_thickness}
\end{gather}
\begin{gather}
  \frac{\mathrm{d}V}{\mathrm{d}t}
  = \frac{1}{M} \left( \frac{U}{H} - \frac{\partial W}{\partial H} \right)
    - \nu \frac{V}{H^2}  \;,
  \label{eq:velocity}
\end{gather} 
where the first term contains the force which is equivalent to the pressure
force and gravity, $\nu$ is viscosity coefficient, the last term corresponds to
the Universe expansion in the standard cosmological model~[37].  Conservation
law for turbulent energy appears in the form
\begin{gather}
  \frac{\mathrm{d}U}{\mathrm{d}t} = - U \frac{V}{H}
    - \epsilon_{\mathrm{d}} \frac{U}{\tau_{\mathrm{d}}}
    + \nu \frac{M V^2}{2 H^2}
    + \epsilon_{\mathrm{SN}} E_{\mathrm{SN}} R_{\mathrm{SN}}  \;.
  \label{eq:internal_energy}
\end{gather}
In this equation the term associated with dissipation (the second term) is
written down taking into account the efficiency of dissipation
$\epsilon_{\mathrm{d}}$ for which an expression was derived in the previous
Section.  We should note that the galactic evolution model that takes into
account dynamic terms similar to~(33) and (34) but does not account for
viscosity, was already derived in~[38].

It is also possible to write down somewhat more accurately the term for
gravitational force that acts upon gaseous disk from disk-like distribution of
the mass in the galaxy:
\begin{multline}
  \frac{\partial W}{\partial H} = \frac{2 G M}{R^2}
    \left( M + M_{\mathrm{s}} \frac{H}{H_{\mathrm{s}}}
      + M_{\mathrm{dm}} \frac{H}{H_{\mathrm{dm}}} \right)  \\
      \nonumber{}\times
    \left( 1 - \frac{H}{\sqrt{R^2 + H^2}} \right)  \;,
\end{multline}
where $M_{\mathrm{s}}$, $H_{\mathrm{s}}$ are, respectively, the mass and
semi-thickness of the stellar disk, $M_{\mathrm{dm}}$, $H_{\mathrm{dm}}$ are
the mass and the length scale of the dark halo distribution.

Viscosity coefficient $\nu$ may be defined as a square of characteristic gas
velocity in the Galaxy, multiplied by the free-fall time:
\begin{multline}
  \nu \sim (100 \text{~km s$^{-1}$})^2\,\tau_{\mathrm{ff}}  \\
  \sim 10^{29}\--10^{30} \text{~cm$^2$ s$^{-1}$}  \;.
\end{multline}
We assumed $\nu = 10^{30}$~cm$^2$ s$^{-1}$ in our computations.

For the star formation law~(10) it is necessary to know the temperature of the
gas.  Let relate turbulent energy and mean temperature in the gas as (see~(9))
\begin{gather}
  U = \frac{3}{2}  \frac{M}{\mu}  k_{\mathrm{B}} T  \;.
  \label{eq:turbulent_energy}
\end{gather}

\section{FORMATION OF THE STELLAR POPULATION OF THE GALAXIES}

Observations of our and external galaxies reveal different populations of
stars: halo stars, stars of the thick and thin disks and bulge.  Galactic
evolution model should describe the origin and properties of different
populations in a uniform way.  Modern models of galactic evolution fall into
two classes~[39]: the first class contains the models of monolithic collapse
with smooth evolutionary transition between stellar populations, while into the
second class belong models in which a galaxy forms by accretion of separate
fragments that experienced an epoch of independent evolution and disk formation
in them is a side effect of accretion. The hierarchical model of formation of
galaxies belongs to the latter class of models.

\begin{figure*}[t!]
  \psfrag{t}{\scriptsize $t$, Gyr}
  \psfrag{lg SFR}{\scriptsize $\lg \text{SFR}$ [$M_\odot$ yr$^{-1}$]}
  \psfrag{lg H}{\scriptsize $\lg H$ [pc]}
  \psfrag{lg (M/M_G)}{\scriptsize $\lg (M/M_\mathrm{G})$}
  \psfrag{lg T}{\scriptsize $\lg T$ [K]}
  \psfrag{V}{\scriptsize $V$, km s$^{-1}$}
  \psfrag{eps_d}{\scriptsize $\epsilon_\mathrm{d}$}
  \psfrag{[O/H]}{\scriptsize $\mathrm{[O/H]}$}
  \psfrag{[Fe/H]}{\scriptsize $\mathrm{[Fe/H]}$}
  \psfrag{[Fe/O]}{\scriptsize $\mathrm{[Fe/O]}$}
  \includegraphics[width=16cm,height=13cm]{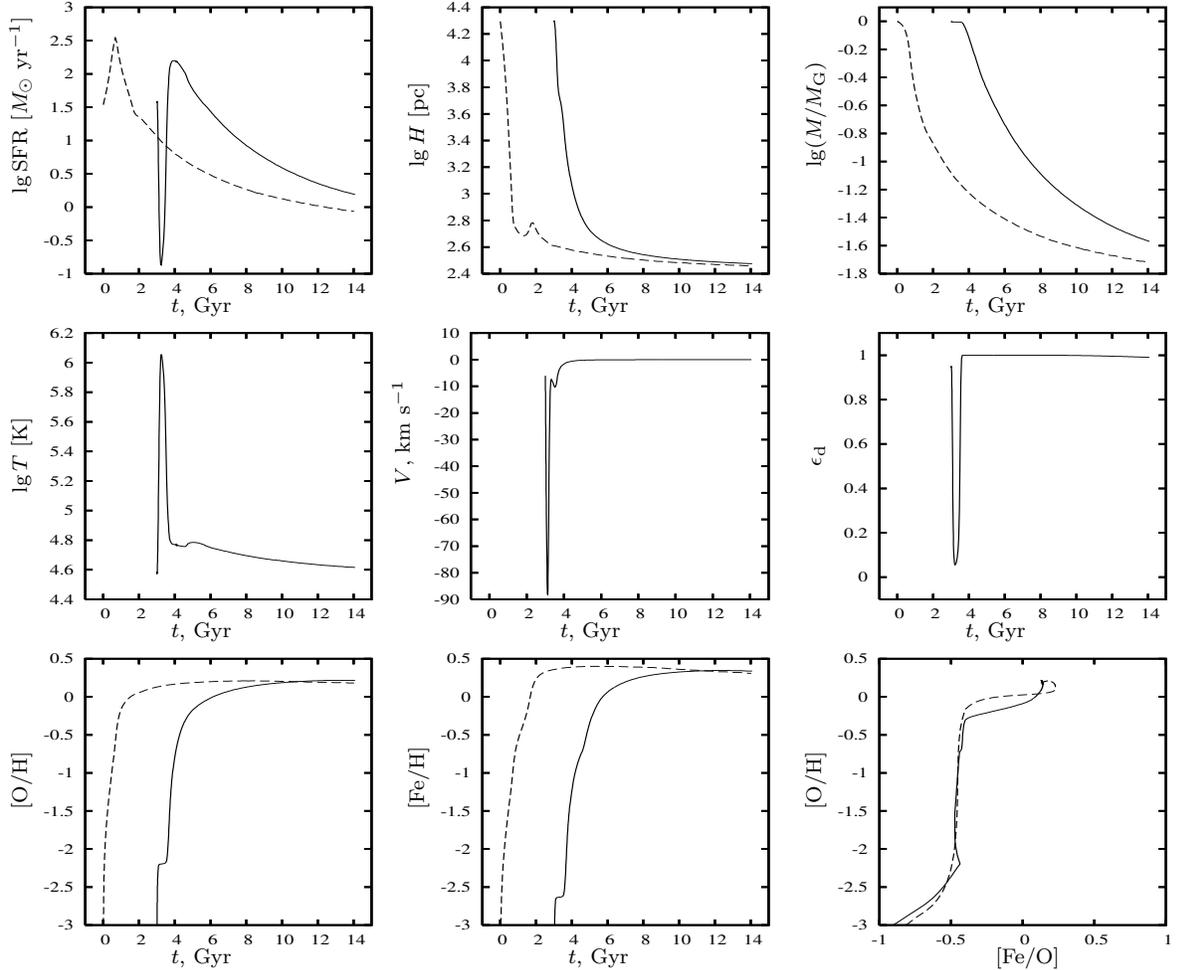}
  \caption{Evolution of the Galaxy in the scenario of monolithic collapse
  (model A); $t$ -- age of the Universe; SFR -- star formation rate; $H$ --
  semi-thickness of the disk; $M/M_{\mathrm{G}}$ -- the ratio of gas mass and
  total mass of the galaxy; $T$ -- temperature of the gas defined by
  formula~(37); $V$ -- velocity of the disk border, $\epsilon_{\mathrm{d}}$
  -- efficiency of dissipation.  Dashed line shows results of computations for
  the standard model~[12].  \hfill}
\end{figure*}

The history of star formation in the galaxies may be strongly non-monotonous
and to reveal several star formation bursts in the lifetime of a galaxy.  Star
formation bursts may be caused, for instance, by gas accretion from
intergalactic medium, by mergers or close passages of the galaxies.  Supernovae
outbursts may delay star formation~[40].  In the star formation history of our
Galaxy one may discover a pause which is seen from the distribution
$\mathrm{[Fe/O]}$~[39] and from dependencies
$\mathrm{[Mg/Fe]}\--\mathrm{[Fe/H]}$~[41] and
$\mathrm{[Eu/Ba]}\--\mathrm{[Fe/H]}$~[42].  This pause corresponds to the range
of stellar ages of $8\--9$ to $10\--12$~Gyr and may be interpreted as an
interruption of the star formation process between the end of formation of
thick disk and the beginning of formation of thin disk~[41, 42].

\begin{figure*}[t!]
  \psfrag{t}{\scriptsize $t$, Gyr}
  \psfrag{lg SFR}{\scriptsize $\lg \text{SFR}$ [$M_\odot$ yr$^{-1}$]}
  \psfrag{lg (M/M_G)}{\scriptsize $\lg (M/M_\mathrm{G})$}
  \psfrag{lg Z}{\scriptsize $\lg Z$}
  \psfrag{[O/H]}{\scriptsize $\mathrm{[O/H]}$}
  \psfrag{[Fe/H]}{\scriptsize $\mathrm{[Fe/H]}$}
  \psfrag{[Fe/O]}{\scriptsize $\mathrm{[Fe/O]}$}
  \includegraphics[width=16cm,height=8.66cm]{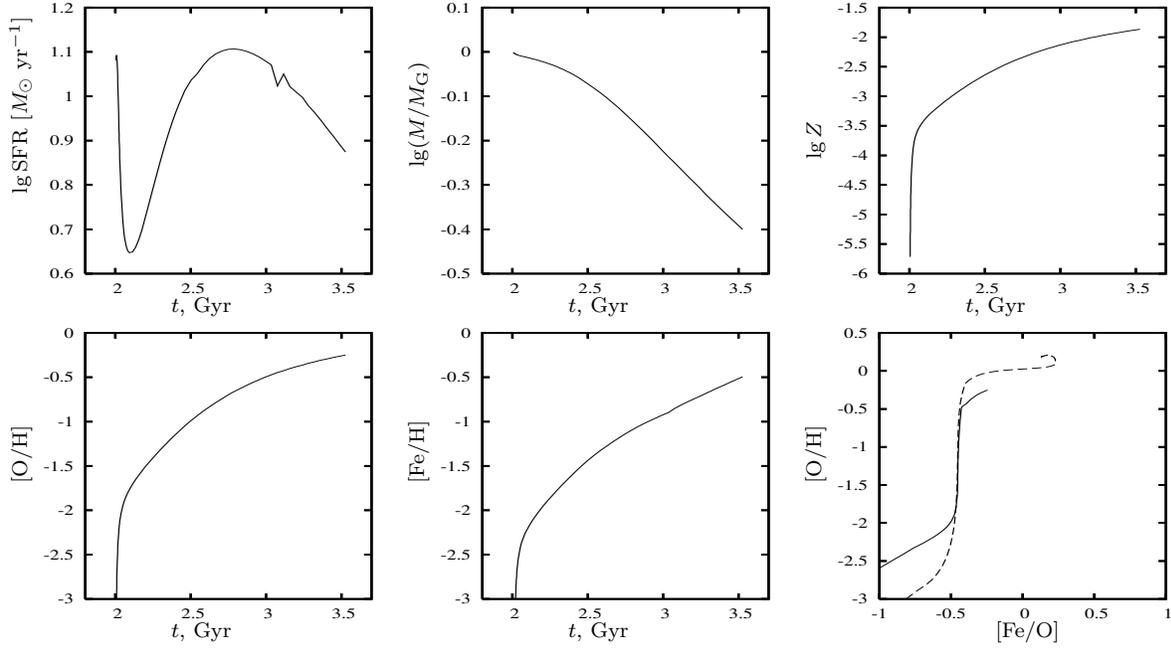}
  \caption{Evolution of a satellite galaxy in the scenario of monolithic
  collapse (model B); $Z$ --- metallicity; the rest of notation as in
  Fig.~1. Dashed line shows results of computations for the standard
  model~[12].  \hfill}
\end{figure*}

\begin{figure*}[t!]
  \psfrag{t}{\scriptsize $t$, Gyr}
  \psfrag{lg SFR}{\scriptsize $\lg \text{óúï}$ ($M_\odot$/ÇÏÄ)}
  \psfrag{lg (M/M_G)}{\scriptsize $\lg (M/M_\mathrm{G})$}
  \psfrag{lg Z}{\scriptsize $\lg Z$}
  \psfrag{[O/H]}{\scriptsize $\mathrm{[O/H]}$}
  \psfrag{[Fe/H]}{\scriptsize $\mathrm{[Fe/H]}$}
  \psfrag{[Fe/O]}{\scriptsize $\mathrm{[Fe/O]}$}
  \includegraphics[width=16cm,height=8.66cm]{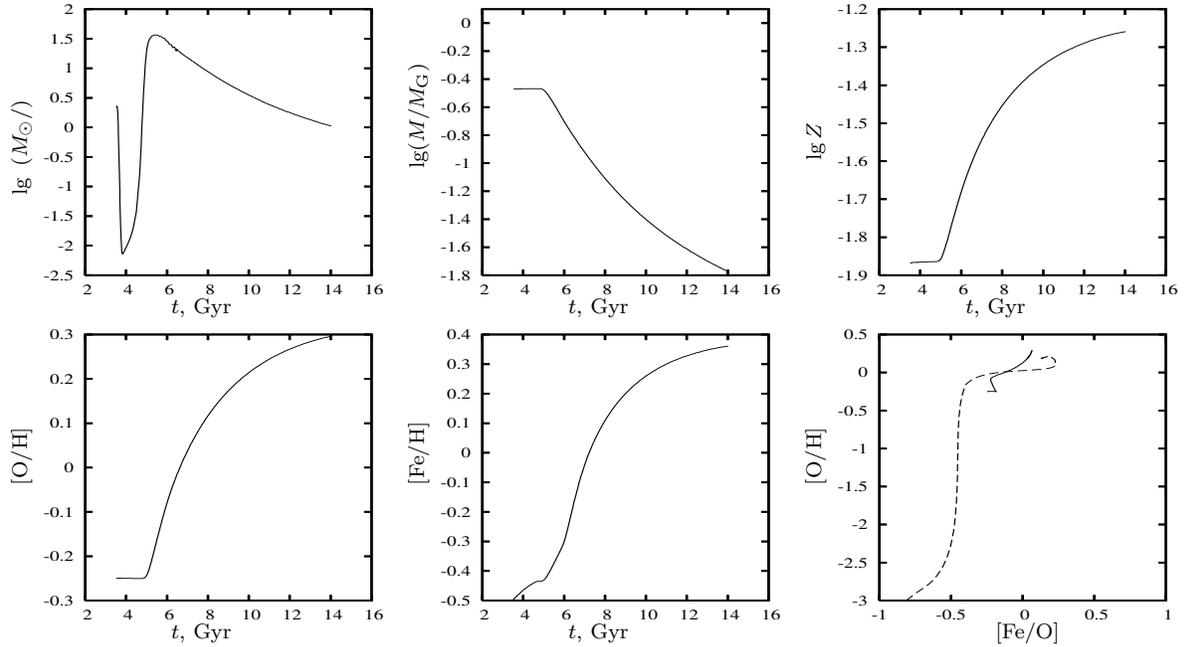}
  \caption{Same as Fig.~2 for the evolution of the Galaxy in the hierarchical
  scenario (model C).  \hfill}
\end{figure*}

The mechanism for cessation of star formation may be exemplified using a closed
(without matter exchange with inter-galactic medium) single-zone model,
appended by equations~(32) and (33), henceforth -- model A.  Figure~1 shows
results of computations for the scenario of monolithic collapse with the
following initial conditions: galactic radius -- $20$~pc, semi-thickness of
protogalactic cloud -- $2\times10^{11}$~$M_{\odot}$, mass of the dark matter
over the length scale of galactic disk is zero, gas temperature --
$4\times10^4$ K, the age of the Universe that corresponds to the beginning of
computations -- $3$~Gyr, ($z = 1.8$ in the standard cosmology~[37]), the rest
of the parameters correspond to standard model~[12].  After the phase of
collapse which results in the increase of gas temperature by an order of
magnitude and a half, star formation interrupts for about $1$~Gyr and then a
burst of star formation occurs with a consequent decay.  As the plots
$\mathrm{[O/H]}\--t$, $\mathrm{[Fe/H]}\--t$, and
$\mathrm{[O/Fe]}\--\mathrm{[Fe/H]}$ show, evolution of chemical abundances also
reflects the presence of the delay of star formation.  However, observations
suggest that iron abundance that corresponds to star formation delay is
$\mathrm{[Fe/H]} \gtrsim -0.5$~[41], while in the model under consideration
cessation of star formation occurs at $\mathrm{[Fe/H]} \lesssim -2.5$.  In the
scenario of monolithic collapse this occurs because of insufficient rate of
star formation at the epoch of initial contraction due to low density of the
protogalactic cloud.

In the computations of Galaxy formation according to the monolithic scenario,
i.e. in model A (Fig.~1), the reason for cessation of star formation is the
increase of temperature, which is due both to gas compression during the
collapse and to the viscous dissipation.  Formation of a shock wave as a result
of collapse or of merging of galaxies may serve as a physical reason for
increase of viscous dissipation.  Larson~[43] has shown that the merging of
galaxies results in the suppression of gas cooling and that this may cause
cessation of star formation. In the hierarchical scenario formation of a galaxy
may occur as a result of a merging of lower-mass galaxies.  Thus, an elliptical
galaxy with mass $\gtrsim 10^{11}$~$M_{\odot}$ may have two or three
significant (i.e., similar mass) precursors and the epoch of its ``assembly''
corresponds to the redshift $z \sim 1$~[44], i.e., to the stellar age of $\sim
9$~Gyr.  Most of the precursors are formed at $z \sim 2\--3$~[44] experiencing
at this time a burst of star formation, during which the abundances of
$\alpha$-elements and heavy elements increase.  The mergings may be important
for disk galaxies too, but in this case the mass of the consumed galaxy has to
be much less, since otherwise an elliptical galaxy will form.  The
metallicities close to solar one that are discovered for some thick disk
stars~[45] also favor hierarchical scenario with independent evolution of
precursors.  More, it is possible to show that the stars of halo and thick disk
formed simultaneously~[42] and formation of halo stars took
$\sim 1.5\times10^9$~yr~[46]; the last circumstance also does not contradict
hierarchical scenario.

\begin{figure*}[t!]
  \includegraphics[width=8cm]{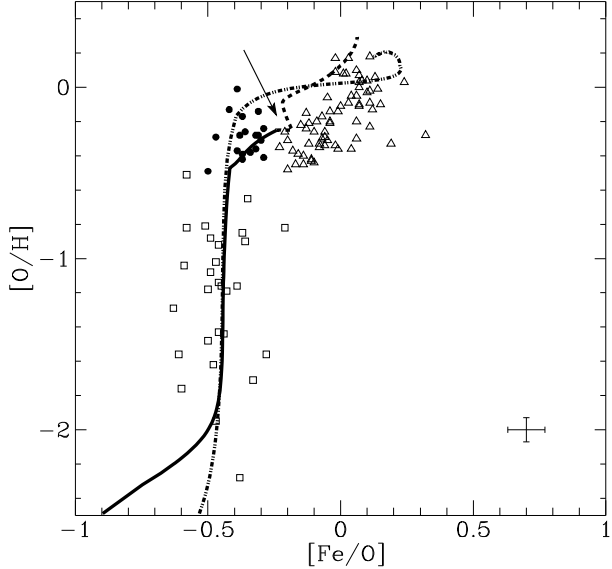}
  \caption{Evolution of the chemical composition of the Galaxy in the
  hierarchical scenario.  We show together the plots of
  $\mathrm{[O/H]}\--\mathrm{[Fe/O]}$ from Figs.~2 and 3.  The arrow indicates
  location of the pause in star formation.  Halo star are shown by squares,
  thick disk stars by circles, and thin disk stars by triangles.  Dashed line
  shows results of computations for the standard model~[12].  \hfill}
\end{figure*} 

In the hierarchical scenario, the history of star formation for our Galaxy that
is consistent with observations may result from a merging with a satellite with
mass $\sim 2\times10^{10}$~$M_\odot$.  Figure~2 shows results of computations
of the evolution of a satellite galaxy with the mass of
$2\times10^{10}$~$M_\odot$, radius $6.32$~kpc, and initial semi-thickness
$6.32$~kpc (these parameters follow Tully-Fisher relation
$M/R^2 \equiv \mathrm{const}$, where $M$ and $R$ are taken from model A); we
designate this model as B. In model B beginning of computations corresponds to
the Universe age $1.5$~Gyr ($z = 2.66$).  It is seen that after the phase of
the initial collapse a burst of star formation occurs which leads to the
increase of oxygen and iron abundances to $\mathrm{[O/H]} \approx -0.25$ and
$\mathrm{[Fe/H]} \approx -0.5$ when the galaxy becomes $1.5$~Gyr old.  One may
infer that merging with satellite occurred at $z = 1.5$ (the age of the
Universe $3.5$~Gyr).  Merging was modeled within scenario of monolithic
collapse (model C) using the following initial conditions: mass of the galaxy
$2.2\times10^{11}$~$M_\odot$, radius -- $20$~kpc, age of the Universe --
$3.5$~Gyr.  At the merging moment, abundance of oxygen in model A was
$\mathrm{[O/H]} \approx -2.2$, abundance of iron --
$\mathrm{[Fe/H]} \approx -2.6$ (Fig.~1), while in model B these abundances were
$\mathrm{[O/H]} \approx -0.25$ and $\mathrm{[Fe/H]} \approx -0.5$, respectively
(Fig.~2). Gas density in our Galaxy was  $0.25$~$M_\odot/$pc$^3$, while in the
satellite it was $0.45$~$M_\odot/$pc$^3$.  This means that in the galactic
region where the merging occurred, mixing of the gas would result in abundances
of oxygen and iron corresponding to the abundances in the satellite galaxy.
This was the reason to set initial chemical composition of the gas similar to
the composition in the model B at the end of computations.

Figure~4 shows relation $\mathrm{[O/H]}\--\mathrm{[Fe/O]}$ composed from the
plots of evolution of abundances of oxygen and iron in models B and C, as well
as positions of halo, thick and thin disk stars according to
Gratton~[39]. Location of the pause in star formation is indicated by the
arrow.  As Fig.~3 shows, the pause in star formation corresponds to the range
of stellar ages $9\--10.5$~Gyr, in agreement with observations.

\section{CONCLUSION}

It is evident that in the star formation law it is necessary to account for the
turbulent energy that results from the turbulization of the ISM by supernovae.
It is also evident that the parameters of the ISM (density, turbulent energy,
temperature) are significantly different at different length scales. In the
present study a star formation function that depends on the turbulent energy of
interstellar medium is suggested.  The model is based on application of Jeans
criterion for power-law distribution of density perturbations in the ISM.  Such
an approach results in the Schmidt-type dependence of the SFR on the gas
density with power $\approx 2$ and an inverse quadratic dependence on turbulent
energy of the ISM (see Eq.~(10)). Applying Jeans criterion, the model may be
generalized to the case when a significant influence of the chemical
composition of the ISM, magnetic field or galactic rotation are assumed.  As
well, we constructed a model for the energy dissipation for the case when the
ISM is considered as turbulent over different length scales.  Star formation
and dissipation laws we suggest, are intended for application in the numerical
models of galactic evolution.  Within single-zone galactic evolution model we
have explained the delay of star formation in the stellar ages range from
$8\--9$ to $10\--12$~Gyr.

\section{ACKNOWLEDGEMENTS}

The author acknowledges B.M.~Shustov, L.I. Mashonkina, D.Z. Wiebe,
Ya.N. Pavlyuchenkov, P.V. Kaigorodov, and A.D. Kudryashov for fruitful
discussions and help in the process of this study.  This study was supported by
the Russian Foundation for Basic Research (project code 05-02-39005-GFEN\_a)
and by State Program of Support of the Leading Scientific Schools of Russian
Federation (project code NSh-4820.2006.2).

\end{document}